\documentclass{iopjournal}

\usepackage{amsmath}
\usepackage{amssymb}
\usepackage{float}
\usepackage{cite}

\begin{document}

\articletype{Paper} 

\title{Operator-Based Information Theory for Imaging: Entropy, Capacity, and Irreversibility in Physical Measurement Systems}

\author{Charles Wood$^1$\orcid{0000-0000-0000-0000}}

\affil{$^1$Future Technology Centre, School of Electrical and Mechanical Engineering, University of Portsmouth, UK}

\email{charles.wood@port.ac.uk}

\keywords{operator theory; information theory; inverse problems; imaging physics; effective rank; irreversibility}

\begin{abstract}
\medskip
Imaging systems are commonly described using resolution, contrast, and signal-to-noise ratio, but these quantities do not provide a general account of how physical transformations affect the flow of information. This paper introduces an operator-based formulation of information theory for imaging.  The approach models the imaging chain as a composition of bounded operators acting on functions, and characterises information redistribution using the spectral properties of these operators. Three measures are developed. Operator entropy quantifies how an operator distributes energy across its singular spectrum. Operator information capacity describes the number of modes that remain recoverable above a noise-dependent threshold. An irreversibility index measures the information lost through suppression or elimination of modes and captures the accumulation of information loss under operator composition. The framework applies to linear, nonlinear, and stochastic operators and does not depend on the specific imaging modality. Analytical examples show how attenuation, blur, and sampling affect entropy, capacity, and irreversibility in different ways. The results provide a general structure for analysing the physical limits of imaging and form the basis for subsequent work on information geometry, spatiotemporal budgets, nonlinear channels, and reconstruction algorithms.
\end{abstract}

\section{Introduction}

\medskip
Imaging systems are often described in terms of resolution, contrast, signal-to-noise ratio, and sampling. These quantities are useful in practice, but they provide only a partial view of how an imaging system transforms information. They sit downstream of a more fundamental question: how does a physical measurement operator redistribute, compress, or destroy the information present in a sample?

\medskip
\noindent Classical information theory treats measurement as a stochastic channel acting on probability distributions~\cite{Shannon1948,CoverThomas2006}. This framework works well for communication problems, but it does not generalise cleanly to physical imaging. Imaging systems act on functions, fields, or volumes through a sequence of physical transformations: illumination, propagation, interaction with the sample, and detection. These processes are naturally expressed as operators~\cite{BarrettMyers2004,BerteroBoccacci1998}. Many of them are non-unitary, nonlinear, or irreversible. Standard channel models cannot capture these behaviours.

\medskip
\noindent This paper develops an operator-based information-theoretic description of physical imaging systems, in which entropy, information capacity, and irreversibility are defined at the level of the measurement operator rather than through data statistics alone. The analysis focuses on the singular value structure of imaging operators and characterises how common physical transformations, such as attenuation, blur, and spectral truncation, redistribute and eliminate information. This formulation makes it possible to identify intrinsic limits on information recovery that are not captured by
stochastic channel models.

\medskip
\noindent Three concepts are developed. The first is \emph{operator entropy}, which characterises how an operator redistributes information across its singular spectrum. The second is \emph{operator information capacity}, defined through the effective rank of the operator and linked to the number of recoverable degrees of freedom. The third is an \emph{irreversibility index}, defined as the fraction of object-space modes that are eliminated (numerically $\sigma_i \le \delta$ on the chosen discretisation) or suppressed below an operating recoverability threshold ($\delta<\sigma_i<\epsilon$).

\medskip
\noindent These definitions allow imaging systems to be analysed on their own terms. They provide a route to quantifying the limits imposed by attenuation, scattering, blur, noise, and sampling, and they extend naturally to nonlinear and physically constrained transformations. Although examples are drawn from X-ray imaging for clarity, the framework is general and applies to any physical measurement system that can be expressed as an operator.

\medskip
\noindent The goal of this work is not to replace existing metrics, but to place them within a broader information-theoretic structure. The operator formulation developed here forms the basis for a general treatment of information flow in imaging, and it prepares the ground for subsequent work on information geometry, spatiotemporal budgets, nonlinear channels, and reconstruction algorithms.

\medskip
\noindent Unlike data-centric measures, the quantities defined here are invariants of the measurement operator (and operating point) and therefore compare physical systems even when signal priors or reconstruction algorithms differ.

\subsection{Why classical information theory is incomplete for physical imaging}
\label{sec:classical_failure}

\medskip
\noindent Classical information theory models measurement as a stochastic channel acting on probability distributions over symbols or random variables. This formulation has been highly successful in communications, but it does not generalise cleanly to physical imaging systems.

\medskip
\noindent Imaging systems do not operate on symbols. They act on functions, fields, or volumes through deterministic physical transformations such as propagation, attenuation, scattering, and detection. These transformations are more naturally expressed as operators acting on function spaces. Information loss in imaging arises not primarily from stochastic uncertainty, but from the
suppression, mixing, or elimination of modes by non-unitary physical operators.

\medskip
\noindent Shannon entropy and mutual information depend on assumed probability distributions over inputs and noise. In imaging, such distributions are often unknown, task-dependent, or imposed for
mathematical convenience rather than physical necessity. As a result, classical information
measures may vary without any corresponding change in the underlying physical measurement
process. More fundamentally, channel-based measures do not distinguish between information that is physically unrecoverable and information that is merely difficult to reconstruct. 

\medskip
\noindent Throughout this paper, \emph{recoverable information} refers to degrees of freedom that can be stably estimated under bounded noise in the induced Hilbert norm. Information is therefore defined operationally in terms of stability of inversion, rather than task-specific decision criteria, priors, or probabilistic loss functions.

\medskip
\noindent Two imaging systems may exhibit similar mutual information or signal-to-noise ratios while differing substantially in which spatial or spectral modes have been eliminated by the measurement
operator. Once such modes are removed by the physics of image formation, no reconstruction
algorithm can restore them.

\medskip
\noindent The operator-based formulation adopted in this work addresses this limitation by shifting the focus from probabilistic descriptions to the spectral structure of the physical measurement operators themselves. Information loss is treated as a property of the operator chain, not of assumed signal or noise statistics. This allows physically imposed limits to be identified independently of priors, reconstruction strategies, or downstream processing.

\medskip
\noindent Task-based, Bayesian, and mutual-information approaches address complementary questions, but they do not isolate operator-induced mode loss independently of priors and reconstruction objectives.

\subsection{Conceptual advance and scope}
\medskip
\noindent The conceptual advance of this work is to shift the description of information loss, capacity, and irreversibility from the level of data and probability distributions to the level of physical measurement operators. In contrast to classical information-theoretic approaches, which characterise information through stochastic models of signals and noise, the present framework treats information flow as an intrinsic property of the operators that implement physical measurement.

\medskip
\noindent In this operator-based view, limits on recoverability are imposed by the structure of the measurement process itself and are independent of reconstruction algorithms, priors, or learning strategies. This separation allows physically irreversible information loss to be identified directly, rather than inferred indirectly from data-level statistics.

\medskip
\noindent The scope of the framework is intentionally general. Although imaging systems motivate the development and examples presented here, the formulation applies to any physical measurement process that can be represented as an operator acting between function spaces. The emphasis of this paper is therefore on establishing a general operator-level description of information flow, rather than on optimising specific modalities or reconstruction methods.

\medskip
\noindent Unlike existing operator-theoretic treatments of imaging, which focus primarily on inversion or stability, the present work introduces entropy, capacity, and irreversibility as intrinsic information measures defined directly on the measurement operator itself.

\section{Imaging as Operator Composition}

\medskip
\noindent Physical imaging systems can be expressed as a sequence of transformations acting on a sample. These transformations are more naturally written as operators on an appropriate function space\cite{BarrettMyers2004,ConwayFA1990}. This section introduces the structure used throughout the paper and defines the classes of operators relevant to imaging.

\subsection{Imaging chains as composed operators}

\medskip
\noindent Let $X$ denote the object function defined on a spatial domain $\Omega \subset \mathbb{R}^n$. An imaging system maps $X$ to a measured image $I$ through a chain of operators. We write this as
\begin{equation}
    I = \mathcal{D} \circ \mathcal{P} \circ \mathcal{S}(X),
\end{equation}
where $\mathcal{S}$ is a source operator, $\mathcal{P}$ is a propagation operator describing the 
interaction between the field and the sample, and $\mathcal{D}$ is the detector response operator. This decomposition is not unique, but it captures the main physical stages of image formation and provides a consistent basis for analysis \cite{BarrettMyers2004,BerteroBoccacci1998}.

\medskip
\noindent \noindent The operators act on functions in $L^2(\Omega)$, the Hilbert space of square-integrable functions on the spatial domain $\Omega$, or on discrete approximations when the continuous model is not required~\cite{ConwayFA1990}. All subsequent definitions apply to either case. The emphasis is on the spectral properties of the operators, not on the specific discretisation.

\subsection{Source operator}

\medskip
\noindent The source operator $\mathcal{S}$ describes how the incident field is shaped before reaching the sample. It may encode spectral filtering, beam shaping, or modulation. In many imaging systems $\mathcal{S}$ is linear, but it is rarely unitary. Absorption, collimation, and spectral weighting 
introduce a non-unitary structure that reduces the information available for downstream recovery.

\subsection{Propagation operator}

\medskip
\noindent The propagation operator $\mathcal{P}$ accounts for the interaction between the incident field and the sample. In scalar formulations this may include attenuation, phase shifts, and scattering. In vector formulations it may also include polarisation-dependent effects. The key property of $\mathcal{P}$ is that it is often non-unitary and, in general, nonlinear. Attenuation acts as a contraction on the input function. Scattering introduces mode mixing and redistribution of energy across spatial frequencies. Dose-dependent changes in the sample introduce nonlinearity. These features make $\mathcal{P}$ a primary source of information loss \cite{BerteroBoccacci1998}.

\subsection{Detector operator}

\medskip
\noindent The detector operator $\mathcal{D}$ describes how the transformed field is mapped onto the image plane. It includes the point-spread function, sampling pattern, noise statistics, and any analogue or digital response curves. In practice $\mathcal{D}$ is a composition of a blur operator, a sampling operator, and a stochastic operator that models noise. The blur component is linear and non-unitary. The sampling component is a projection onto a discrete grid. The noise component introduces stochasticity but can, in principle, be treated within an operator framework when considered as a conditional expectation or an averaging operator \cite{BerteroBoccacci1998}.

\subsection{Classes of operators in imaging}

\medskip
\noindent We distinguish three classes of operators that occur frequently in imaging:
\begin{enumerate}
    \item \textbf{Linear, non-unitary operators}. These include attenuation, convolutional blur, and 
          sampling. Their singular value spectra directly determine how information is redistributed.
    \item \textbf{Nonlinear operators}. These arise from dose-dependent changes in the sample, 
          nonlinear detector responses, and certain scattering processes. Their information properties 
          can be analysed through local linearisation.
    \item \textbf{Stochastic operators}. Noise processes can be represented as conditional expectation 
          operators, which act as contractions on the space of possible signals.
\end{enumerate}

\noindent All physically real imaging systems are compositions of these classes. This motivates an operator-based information theory: information loss, redistribution, and irreversibility arise from the spectral and geometric properties of the composed operator.

\subsection{Motivation for an operator framework}

\medskip
\noindent Conventional descriptions of imaging quantify performance using resolution, contrast, and noise. These metrics depend on the specific form of the composed operator but do not characterise it directly. By treating the imaging chain as a single operator, or as a sequence of operators with known spectral structure, we can analyse information flow without committing to a particular modality or reconstruction method. This provides a general foundation for the entropy and capacity measures developed in the next 
section.

\begin{table}[H]
\centering
\begin{tabular}{ll}
\hline
Symbol & Meaning \\
\hline
$X$ & Object function defined on a spatial domain $\Omega$ \\
$I$ & Measured image \\
$\mathcal{S}$ & Source operator \\
$\mathcal{P}$ & Propagation operator \\
$\mathcal{D}$ & Detector operator \\
$\mathcal{O}$ & Generic operator acting on a function space \\
$\sigma_i$ & Singular values of $\mathcal{O}$ \\
$\lambda_i$ & Normalised singular weights, $\lambda_i = \sigma_i^2 / \sum_j \sigma_j^2$ \\
$H(\mathcal{O})$ & Operator entropy \\
$C(\mathcal{O})$ & Operator information capacity \\
$\mathcal{I}_\epsilon(\mathcal{O})$ & Irreversibility index at operating threshold $\epsilon$ \\
$\mathrm{rank}_\epsilon(\mathcal{O})$ & Effective rank above threshold $\epsilon$ \\
$\delta$ & Numerical/modelling tolerance used to define ``numerically zero'' singular values \\
$\circ$ & Composition of operators \\
\hline
\end{tabular}
\caption{Notation used throughout the paper.}
\end{table}

\section{Operator Entropy}

\medskip
\noindent The purpose of this section is to define a measure of entropy for operators that act on functions or discrete signals in imaging. The aim is to quantify how an operator redistributes information across its singular spectrum. The motivation is straightforward: imaging operators reduce, compress, or eliminate modes of the input function, and the singular values provide a direct way to measure these effects.

\medskip
\noindent\textbf{Well-definedness and discretisation.}
Throughout this paper, operator spectra are interpreted either (i) on a finite-dimensional discretisation of the object space induced by a chosen basis and reconstruction grid, or (ii) in the infinite-dimensional setting for compact operators with square-summable singular values (Hilbert--Schmidt class). In case (i), all spectral sums are finite. In case (ii), normalisation by $\sum_i \sigma_i^2$ is well-defined whenever $\{\sigma_i\}\in \ell^2$ (square-summable), and the resulting quantities are basis-invariant within the chosen Hilbert space structure. The analysis below is therefore operationally tied to the discretisation used in reconstruction, while remaining consistent with standard operator-theoretic conditions where infinite-dimensional limits are taken.

\medskip
\noindent In what follows, all spectral quantities are therefore understood either on a finite discretisation (always well-defined) or for compact/Hilbert--Schmidt operators where $\sum_i \sigma_i^2 < \infty$.

\subsection{Motivation}

\medskip
\noindent Classical entropy measures describe uncertainty in a probability distribution~\cite{Shannon1948,CoverThomas2006}. Imaging operators do not act on probability distributions; they act on functions. Their information properties depend on how they transform the energy of different spatial modes. For a bounded linear operator $\mathcal{O}: \mathcal{H} \rightarrow \mathcal{H}$ on a Hilbert space $\mathcal{H}$ (e.g.\ $L^2(\Omega)$ or its finite-dimensional discretisation), the singular 
value decomposition expresses $\mathcal{O}$ as
\begin{equation}
    \mathcal{O} = U \, \Sigma \, V^\ast,
\end{equation}
where $\Sigma$ contains the singular values $\sigma_i \ge 0$~\cite{ConwayFA1990}. The singular values specify how much 
each input mode contributes to the output. Modes with singular values that are zero (or numerically zero under the chosen tolerance) are eliminated. This motivates defining entropy directly from the normalised singular values.

\subsection{Definition}

\medskip 
Let $\{\sigma_i\}$ be the singular values of $\mathcal{O}$. We define normalised weights
\begin{equation}
    \lambda_i = \frac{\sigma_i^2}{\sum_j \sigma_j^2},
\end{equation}

\noindent which describe the relative contribution of each mode to the output energy. 

\medskip
\noindent The squared singular values $\sigma_i^2$ weight the transmitted energy in mode $i$ under unit-norm excitation in the corresponding singular vector basis. Normalisation by $\sum_j \sigma_j^2=\|\mathcal{O}\|_{HS}^2$ (or its discrete analogue) therefore yields a physically interpretable modal energy distribution. 

\medskip
\noindent Other spectral weightings could be adopted to emphasise different notions of transmission, but $\sigma_i^2$ is the most direct choice for energy-based operator normalisation in imaging.

\medskip
\noindent The \emph{operator entropy} 
is defined as
\begin{equation}
    H(\mathcal{O}) = - \sum_i \lambda_i \log \lambda_i.
\end{equation}

\medskip
\noindent An equivalent ``effective number of modes'' associated with $\mathcal{O}$ is given by
\begin{equation}
    r_{\mathrm{eff}}(\mathcal{O}) = \exp\!\big(H(\mathcal{O})\big),
\end{equation}
which connects operator entropy directly to the effective-rank literature.

\medskip
\noindent This definition satisfies the basic requirements expected of an entropy measure:
\begin{enumerate}
    \item $H(\mathcal{O}) = 0$ when the operator retains energy in a single mode.
    \item $H(\mathcal{O})$ is maximal when all modes contribute equally.
    \item $H(\mathcal{O})$ decreases as the singular spectrum becomes more concentrated.
\end{enumerate}

\noindent The measure is scale-invariant because the singular values are normalised. Global scaling of the 
operator does not change the entropy.

\subsection{Interpretation}

\medskip
\noindent The operator entropy quantifies the diversity of modes preserved by $\mathcal{O}$. A high-entropy operator spreads information across many modes. A low-entropy operator concentrates energy into few modes or eliminates modes entirely.

\medskip
\noindent In imaging:
\begin{itemize}
    \item uniform attenuation is a scalar contraction and leaves $H(\mathcal{O})$ unchanged (it scales all singular values equally), but it can reduce $C(\mathcal{O})$ by pushing modes below the recoverability threshold; spatially varying attenuation may alter relative modal weights and can change $H(\mathcal{O})$;
    \item blur operators reduce entropy by suppressing fine-scale structure;
    \item sampling reduces entropy by projecting onto a lower-dimensional subspace;
    \item scattering can either increase or decrease entropy depending on whether it redistributes 
          energy across modes or eliminates them.
\end{itemize}

\noindent Operators with exact (or numerical) rank deficiency have lower entropy than full-rank operators on the same discretised space. This reflects the irretrievable loss of modes.

\subsection{Relation to classical entropy}

\medskip
\noindent When $\mathcal{O}$ is a convolution operator with a known transfer function, the singular values correspond to squared magnitudes of the transfer function. In this case the operator entropy reduces to a measure of frequency-domain spreading. This recovers a form of Shannon’s entropy for linear channels~\cite{Shannon1948} but extends naturally to non-unitary operators that are not modelled as stochastic channels. Similar entropy measures based on singular values exist in the effective-rank literature~\cite{RoyVetterli2007} 
and in threshold-based spectral analysis~\cite{GavishDonoho2014}.

\medskip
\noindent The operator entropy, therefore, provides a general way of quantifying information redistribution in deterministic, physically-constrained transformations without resorting to probability distributions.

\subsection{Role in imaging}

\medskip
\noindent The operator entropy will be used in later sections to define:
\begin{enumerate}
    \item the information capacity of an operator, 
    \item the irreversibility of operator composition,
    \item the information limits of imaging chains.
\end{enumerate}

\noindent It provides a quantitative link between the physical action of an imaging system and the information that can be recovered downstream.

\section{Operator Information Capacity}

\medskip
\noindent The operator entropy defined in the previous section quantifies how an operator redistributes information across its singular spectrum. Entropy alone does not specify how many independent degrees of freedom can be recovered from the output. For this purpose we require a measure of information capacity that reflects the effective number of modes preserved by the operator. This section defines such a measure and establishes its basic properties.

\subsection{Definition}

\medskip
Let $\mathcal{O}$ be a bounded linear operator with singular values $\{\sigma_i\}$. In any 
physical imaging system, modes with singular values below a noise threshold cannot be recovered 
reliably. This motivates defining the \emph{effective rank} as~\cite{RoyVetterli2007}
\begin{equation}
    \mathrm{rank}_{\epsilon}(\mathcal{O}) = 
    \# \{\, i : \sigma_i \ge \epsilon \,\},
\end{equation}
where $\epsilon$ is a noise- and task-dependent recoverability threshold. We assume $\epsilon \ge \delta$, since modes below the numerical or modelling tolerance are unrecoverable regardless of noise level. This ensures that the recoverability threshold exceeds the tolerance used to define effective nullspace modes.

\medskip
\noindent To make this explicit, consider a standard measurement model
\begin{equation}
    y = \mathcal{O}x + n,
\end{equation}
where $n$ represents additive detector noise with $\mathbb{E}\|n\|^2 = \sigma_n^2$ (or an equivalent noise bound). 

\medskip
\noindent In the singular-vector basis of $\mathcal{O}$, the $i$th measured coefficient scales as $\sigma_i x_i$, where $x_i$ is the corresponding input coefficient, and is recoverable only if it exceeds the noise floor by a chosen criterion, for example
\begin{equation}
    \sigma_i |x_i| \ge \kappa \sigma_n,
\end{equation}
where $\kappa$ is set by the desired reconstruction tolerance or error probability. 

\medskip
\noindent This motivates a recoverability condition of the form $\sigma_i \ge \kappa \sigma_n / |x_i|$. In order to define a \emph{single} operating-point threshold $\epsilon$ for $\mathrm{rank}_\epsilon(\mathcal{O})$, we take $\epsilon$ to be the resulting \emph{spectral} threshold obtained by fixing a conservative or typical amplitude scale for the object space (e.g.\ via $\|x\|$ bounds, calibration phantoms, or application-specific dynamic range), so that $\sigma_i \ge \epsilon$ encodes stable recoverability under the chosen noise tolerance.

\medskip
\noindent In practice, $\epsilon$ is treated as the system operating-point threshold implied by detector noise, exposure, and reconstruction tolerance; principled choices of such spectral thresholds are discussed in the effective-rank and optimal-threshold literature~\cite{GavishDonoho2014,RoyVetterli2007}. Related notions of numerical rank, spectral truncation, and stability in discrete inverse problems are treated extensively in the inverse problems literature~\cite{Hansen1998}.

\medskip
\noindent The \emph{operator information capacity} is then
\begin{equation}
    C(\mathcal{O}) = \log \big( \mathrm{rank}_{\epsilon}(\mathcal{O}) \big).
\end{equation}

\medskip
\noindent The term ``capacity'' is used here in an operator-theoretic sense, distinct from Shannon capacity. It characterises recoverable dimensionality under physical and noise constraints, rather than achievable bit rates under coding assumptions. This definition reflects the number of independent modes that remain above the noise floor. The logarithm aligns the measure with standard information-theoretic quantities and provides a scale that grows additively when independent channels are combined.

\subsection{Interpretation}

\medskip The capacity $C(\mathcal{O})$ characterises the amount of information that can, in principle, be recovered after the action of the operator. If $\mathcal{O}$ has full numerical rank then $C(\mathcal{O})$ is maximal. If attenuation, blur, or sampling suppress a subset of modes below $\epsilon$, the capacity decreases accordingly.

\medskip
\noindent In imaging this quantity has a direct interpretation. It determines:
\begin{itemize}
    \item the number of recoverable spatial modes,
    \item the effective dimensionality of the reconstruction problem,
    \item the best achievable resolution under noise,
    \item the stability of inverse problems associated with $\mathcal{O}$.
\end{itemize}

\noindent A reduction in capacity cannot be restored by any post-processing. Once modes fall below the noise threshold they are effectively lost.

\subsection{Nonlinear operators}

\medskip
\noindent Many physical imaging operators are nonlinear. For a differentiable nonlinear operator 
$\mathcal{O}: X \mapsto \mathcal{O}(X)$, the local behaviour is governed by the Fréchet derivative $D\mathcal{O}(X)$~\cite{Engl1996}. The local capacity is defined as
\begin{equation}
    C_X(\mathcal{O}) = C\big(D\mathcal{O}(X)\big).
\end{equation}
This captures how many modes can be recovered in a neighbourhood of $X$. For operators that depend 
on illumination dose, scattering strength, or detector saturation, $C_X(\mathcal{O})$ provides a 
direct measure of the information preserved at that operating point.

\subsection{Relation to operator entropy}

\medskip
\noindent Capacity and entropy provide complementary information. Capacity specifies how many modes are recoverable. Entropy specifies how the operator distributes energy across those modes. An operator may have high entropy, but low capacity if it spreads energy across many modes while suppressing a large subset below the noise threshold. Conversely, an operator may have low entropy, but high capacity if it preserves a small set of modes with similar strength. This distinction becomes important when analysing irreversibility in the next section.

\subsection{Role in imaging}

\medskip
\noindent The operator information capacity provides a general criterion for comparing imaging configurations. Increasing source coherence, improving detector sampling, or reducing propagation loss all act to increase $\mathrm{rank}_{\epsilon}(\mathcal{O})$ and therefore raise $C(\mathcal{O})$. 

\medskip
\noindent Conversely, any physical mechanism that eliminates modes or pushes them below the noise floor reduces capacity. 

\medskip
\noindent The measure also provides a consistent framework for analysing multi-stage imaging chains. For operators $\mathcal{A}$ and $\mathcal{B}$ acting on the same discretised object space, singular values obey the bound
\begin{equation}
    \sigma_k(\mathcal{B}\circ\mathcal{A}) \le \|\mathcal{B}\|\,\sigma_k(\mathcal{A}),
\end{equation}
and similarly $\sigma_k(\mathcal{B}\circ\mathcal{A}) \le \|\mathcal{A}\|\,\sigma_k(\mathcal{B})$. Consequently, the $\epsilon$-effective rank satisfies
\begin{equation}
    \mathrm{rank}_{\epsilon}(\mathcal{B}\circ\mathcal{A})
    \le \min\!\big(\mathrm{rank}_{\epsilon/\|\mathcal{B}\|}(\mathcal{A}),\,
                   \mathrm{rank}_{\epsilon/\|\mathcal{A}\|}(\mathcal{B})\big),
\end{equation}
which captures the fact that recoverability cannot increase under composition without a change in operating point or norming.

\medskip
\noindent Capacity therefore provides a quantitative link between operator structure and the limits of reconstruction.

\section{Irreversibility of Operator Composition}

\medskip
\noindent The previous sections introduced operator entropy and operator information capacity as
measures of how a transformation redistributes and preserves information. This section defines a
measure of irreversibility that quantifies how many degrees of freedom are removed from the object
space by the physical measurement operator at a given operating threshold. The motivation is
straightforward: imaging operators are rarely invertible, and information lost at any stage cannot
be restored downstream.

\subsection{Definition}

\medskip
\noindent Let $\mathcal{O}$ be a bounded linear operator with singular values $\{\sigma_i\}$ on a
fixed discretised object space of dimension $N$ (or an $N$-mode truncation for compact operators).

\medskip
\noindent Let $\delta>0$ denote a numerical or modelling tolerance (e.g.\ machine precision and/or forward-model error) used to define ``numerically zero'' singular values on the chosen discretisation.

\medskip
\noindent We define the set of \emph{recoverable modes} at operating threshold $\epsilon$ as
\begin{equation}
    \mathcal{R}_\epsilon(\mathcal{O}) = \{\, i : \sigma_i \ge \epsilon \,\},
\end{equation}
and the set of \emph{lost modes} as
\begin{equation}
    \mathcal{L}_\epsilon(\mathcal{O}) = \{\, i : \sigma_i \le \delta \,\} \cup
    \{\, i : \delta < \sigma_i < \epsilon \,\}.
\end{equation}

\medskip
\noindent The \emph{irreversibility index} is then defined as the fraction of modes that are lost:
\begin{equation}
    \mathcal{I}_\epsilon(\mathcal{O}) = \frac{|\mathcal{L}_\epsilon(\mathcal{O})|}{N}
    = 1 - \frac{|\mathcal{R}_\epsilon(\mathcal{O})|}{N}.
\end{equation}

\noindent This definition separates two physically distinct mechanisms: modes with singular values
that are \emph{numerically zero on the chosen discretisation} (i.e.\ $\sigma_i \le \delta$ under the chosen numerical or modelling tolerance) are eliminated in the discretised forward model and constitute \emph{hard loss} (numerical nullspace loss), while modes with $\delta < \sigma_i < \epsilon$ are present but unrecoverable at the operating point and constitute \emph{soft loss}. Both contribute to irreversibility in the sense of unrecoverable degrees of freedom from measured data at the chosen noise tolerance. 

\medskip
\noindent Hard loss may arise from true physical nullspace and/or from discretisation or forward-model error; the framework treats it operationally as numerical nullspace loss at tolerance $\delta$.

\medskip
\noindent The distinction between hard and soft loss is therefore defined relative to the discretised forward model used for reconstruction, consistent with standard notions of numerical rank in inverse problems.

\subsection{Interpretation}

\medskip
\noindent The irreversibility index quantifies the fundamental loss introduced by an operator. If
attenuation or blur suppress certain modes below the recoverable threshold, those modes contribute
to $\mathcal{I}_\epsilon(\mathcal{O})$. If an operator eliminates modes entirely, the index
increases further.

\medskip
\noindent In physical imaging, irreversibility is introduced by:
\begin{itemize}
    \item attenuation and dose limits, which act as contractions and can push modes below the recoverability threshold;
    \item blur, which suppresses fine structure below the noise floor;
    \item sampling, which projects onto a lower-dimensional subspace;
    \item noise, which effectively raises the threshold $\epsilon$.
\end{itemize}

\noindent These mechanisms act independently but have a cumulative effect on irreversibility.

\subsection{Composition rule}

\medskip
\noindent For two operators $\mathcal{A}$ and $\mathcal{B}$ acting on the same discretised object
space of dimension $N$, singular values obey the bound
\begin{equation}
    \sigma_k(\mathcal{B}\circ\mathcal{A}) \le \|\mathcal{B}\|\,\sigma_k(\mathcal{A}),
\end{equation}
and similarly $\sigma_k(\mathcal{B}\circ\mathcal{A}) \le \|\mathcal{A}\|\,\sigma_k(\mathcal{B})$.
Consequently, the $\epsilon$-effective rank satisfies
\begin{equation}
    \mathrm{rank}_{\epsilon}(\mathcal{B}\circ\mathcal{A})
    \le \min\!\big(\mathrm{rank}_{\epsilon/\|\mathcal{B}\|}(\mathcal{A}),\,
                   \mathrm{rank}_{\epsilon/\|\mathcal{A}\|}(\mathcal{B})\big).
\end{equation}

\medskip
\noindent Since $\mathcal{I}_\epsilon(\mathcal{O}) = 1 - \mathrm{rank}_{\epsilon}(\mathcal{O})/N$ on a fixed
object space, this implies a corresponding monotonicity statement with norm-adjusted thresholds. In the common non-expansive case (after normalisation) where $\|\mathcal{A}\|\le 1$ and $\|\mathcal{B}\|\le 1$, recoverability cannot increase at fixed $\epsilon$ on the same discretised object space, and thus $\mathcal{I}_\epsilon$ is non-decreasing under composition at that operating point. This statement concerns recoverability at a fixed operating point and does not assert monotonicity of individual singular vectors, which may rotate under composition.

\subsection{Role in imaging}

\medskip
\noindent The irreversibility index provides a direct way to compare imaging configurations.
Improving source coherence, reducing propagation loss, or increasing detector sampling can reduce
$\mathcal{I}_\epsilon(\mathcal{O})$. Conversely, stronger attenuation, increased blur, or
undersampling increase irreversibility. The index also clarifies the limits of post-processing. No
reconstruction algorithm can reduce $\mathcal{I}_\epsilon(\mathcal{O})$ at a fixed operating
threshold. It is set entirely by the physical transformations applied to the sample. Any downstream
processing is constrained by this value.

\subsection{Irreversibility as a physical invariant of imaging chains}
\label{sec:irreversibility_invariant}

\medskip
\noindent The irreversibility index defined in this work is not a property of reconstruction,
regularisation, or algorithmic sophistication. It is a property of the physical operator chain
itself. Once a mode is eliminated (that is, lies in the nullspace of the measurement operator), it
is not identifiable from the measured data by any inverse or data-driven reconstruction method
without introducing additional external information. Modes suppressed below the recoverability
threshold may be estimated only through the imposition of priors, regularisation, or learned
structure, which necessarily introduce bias and do not increase the number of recoverable degrees
of freedom supported by the physical operator. Reconstruction methods can, therefore, reorganise
information within the surviving subspace, but cannot expand the effective range of the measurement
operator.

\medskip
\noindent In this sense, $\mathcal{I}_\epsilon(\mathcal{O})$ is an \emph{operator-level invariant} of the imaging chain at a fixed discretisation and operating point: for a fixed physical operator $\mathcal{O}$, fixed discretised object space, and fixed operating threshold $\epsilon$, its value is unaffected by reconstruction algorithms, priors, or post-processing. It changes only when the measurement operator or operating point changes (e.g., through sampling, optics, exposure, detector noise, or a different choice of $\epsilon$). This distinction clarifies the boundary between physical and computational limits in imaging: algorithms can only reorganise estimates within the recoverable subspace implied by $\mathcal{O}$ at $\epsilon$, and cannot reduce $\mathcal{I}_\epsilon(\mathcal{O})$ without altering the physics or the operating point.

\section{Analytical Examples}

\medskip
\noindent This section illustrates the behaviour of operator entropy, operator information capacity, and irreversibility using three analytical examples. These examples are chosen because they represent common transformations in imaging and have singular value spectra that can be described explicitly. No experimental data are required.

\subsection{Attenuation operator}

\medskip
\noindent Consider a multiplicative attenuation operator acting on a function $x(\mathbf{r})$,
\begin{equation}
    \mathcal{A}(x)(\mathbf{r}) = e^{-\mu(\mathbf{r}) d} \, x(\mathbf{r}),
\end{equation}
where $\mu(\mathbf{r})$ is an attenuation coefficient and $d$ is the propagation distance. When 
$\mu$ is constant, $\mathcal{A}$ is a scalar contraction with singular values
\begin{equation}
    \sigma_i = e^{-\mu d}.
\end{equation}

\noindent All singular values are equal, so the normalised weights satisfy $\lambda_i = 1/N$ for an $N$-dimensional discretisation. 

\medskip
\noindent\textbf{Remark (spatially varying attenuation).}
When $\mu(\mathbf{r})$ varies spatially, $\mathcal{A}$ is still a multiplication operator in the spatial , but is no longer a scalar multiple of the identity on a generic discretisation. In particular, relative modal weights may change under a chosen basis, and operator entropy need not remain maximal. The constant-$\mu$ case is presented here only to isolate pure contraction without spectral redistribution.

\medskip
\noindent The entropy is therefore
\begin{equation}
    H(\mathcal{A}) = \log N,
\end{equation}
which reflects the fact that the operator preserves the relative weighting of all modes. However, the 
capacity is reduced if $e^{-\mu d} < \epsilon$, in which case all modes fall below the noise 
threshold and
\begin{equation}
    C(\mathcal{A}) = 0.
\end{equation}
This example shows that a contraction can preserve entropy while eliminating recoverable modes.

\subsection{Gaussian blur operator}

\medskip
Let $\mathcal{B}$ be convolution with a Gaussian kernel,
\begin{equation}
    \mathcal{B}(x) = g_\sigma \ast x,
\end{equation}
where $g_\sigma$ is the Gaussian of width $\sigma$. In the Fourier basis the singular values are
\begin{equation}
    \sigma(\mathbf{k}) = e^{-\frac{1}{2}\sigma^2 \|\mathbf{k}\|^2}.
\end{equation}
High-frequency modes are suppressed exponentially as $\|\mathbf{k}\|$ increases.

\medskip
\noindent The operator entropy depends on the spread of $\sigma(\mathbf{k})^2$. As $\sigma$ increases, the spectral distribution becomes increasingly concentrated at low frequencies, reducing $H(\mathcal{B})$. The quantitative values of entropy and capacity depend on discretisation, bandwidth, and truncation, but the qualitative trend reflects progressive suppression of high-frequency modes by the blur operator. The capacity decreases as more modes fall below $\epsilon$, and the irreversibility index increases correspondingly. This operator demonstrates how blur reduces both entropy and recoverable dimensionality.

\subsection{Sampling operator}

\medskip Let $\mathcal{S}$ be an orthogonal projection onto a subset of sample points. In a discrete setting, if $M$ out of $N$ points are sampled then $\mathcal{S}$ has singular values
\begin{equation}
    \sigma_i = 
    \begin{cases}
        1, & i \le M, \\
        0, & i > M.
    \end{cases}
\end{equation}
The operator entropy is
\begin{equation}
    H(\mathcal{S}) = \log M,
\end{equation}
reflecting the fact that $M$ modes are preserved with equal weight. For any operating threshold $\epsilon \le 1$, the capacity is
\begin{equation}
    C(\mathcal{S}) = \log M.
\end{equation}

\noindent The irreversibility index is defined relative to the pre-measurement object space. On the original
$N$-dimensional object space, the sampling operator eliminates $N-M$ modes exactly, giving
\begin{equation}
    \mathcal{I}_\epsilon(\mathcal{S}) = \frac{N-M}{N}.
\end{equation}
Restricting the object space to the $M$-dimensional sampled subspace yields zero irreversibility by
construction, but removes the eliminated modes from consideration and therefore does not quantify
information loss introduced by the measurement. In this work, irreversibility is always evaluated
on the original object space.

\subsection{Distinct mechanisms of information loss}
\label{sec:loss_mechanisms}

\medskip
\noindent The analytical examples above highlight three distinct mechanisms by which information is lost in physical imaging systems:
\begin{enumerate}
    \item contraction without spectral redistribution (attenuation),
    \item progressive suppression of high-frequency modes (blur),
    \item hard elimination of modes through projection (sampling).
\end{enumerate}

\noindent Although these mechanisms all reduce the information available for reconstruction, they do so in fundamentally different ways. Attenuation preserves the relative weighting of modes while reducing their absolute magnitude. Blur redistributes energy toward lower frequencies, progressively suppressing fine-scale structure. Sampling removes modes entirely by projecting onto a lower-dimensional subspace.

\medskip
\noindent Conventional imaging metrics such as resolution, contrast, and signal-to-noise ratio do not distinguish reliably between these mechanisms. As a result, physically inequivalent imaging systems may appear equivalent under standard performance measures. The consequences of this limitation are illustrated explicitly in the following comparative example.

\subsection{Comparative example: metric equivalence versus operator inequivalence}
\label{sec:comparative_example}

\medskip
\noindent To illustrate the limitations of conventional imaging metrics, consider two linear imaging operators, $\mathcal{O}_A$ and $\mathcal{O}_B$, acting on the same object space.

\medskip
\noindent Operator $\mathcal{O}_A$ consists of a mild Gaussian blur followed by coarse sampling. Operator $\mathcal{O}_B$ consists of stronger Gaussian blur followed by denser sampling. The parameters are chosen such that both systems exhibit comparable nominal resolution, similar modulation transfer function cut-offs, and identical signal-to-noise ratios at the detector.

\medskip 
\noindent Under conventional assessment, these systems would be judged equivalent. Resolution metrics report similar spatial resolving power, signal-to-noise ratios are matched by construction, and contrast transfer appears comparable within the passband.

\medskip
\noindent However, the singular value spectra of the two operators differ substantially. In $\mathcal{O}_A$, high-frequency modes are sharply truncated by sampling, resulting in hard rank deficiency. In $\mathcal{O}_B$, modes are progressively attenuated yet remain present above the noise threshold across a broader spectral range.

\medskip
\noindent As a consequence, the operator information capacity satisfies
\begin{equation}
    C(\mathcal{O}_B) > C(\mathcal{O}_A),
\end{equation}
despite the apparent equivalence of the two systems under conventional metrics. 

\medskip
\noindent The operator
entropy also differs, reflecting distinct patterns of spectral redistribution. Most importantly,
the irreversibility index of $\mathcal{O}_A$ exceeds that of $\mathcal{O}_B$, indicating that a larger
fraction of information has been eliminated entirely rather than suppressed.

\medskip
\noindent This example demonstrates that resolution, contrast, and signal-to-noise ratio are insufficient to characterise information preservation in imaging systems. The operator-based measures introduced here distinguish between suppression and elimination of modes, and reveal physically meaningful differences that conventional metrics cannot capture.

\subsection{Limitations of data-centric information measures}

\medskip
\noindent The comparative example above highlights a limitation of conventional, data-centric information measures. Quantities such as Shannon entropy~\cite{Shannon1948} characterise the statistical structure of measured data, but do not encode how that structure arises from the measurement process itself. In particular, they do not distinguish between information preserved through an invertible operator and information eliminated by a non-invertible transformation.

\medskip
\noindent As demonstrated in Section~\ref{sec:comparative_example}, two imaging systems may produce data with similar resolution, signal-to-noise ratio, and even comparable data-level entropy, while differing substantially in their ability to preserve recoverable information. In the case of hard truncation, entire subspaces of the object are removed by the operator, whereas progressive attenuation suppresses modes without fully eliminating them. From a data-centric perspective, these mechanisms may appear equivalent.

\medskip
\noindent This inability to distinguish suppression from elimination reflects a fundamental limitation of information measures that are defined solely on observed data. By assigning informational meaning directly to the measurement operator, the operator-based framework introduced here captures irreversibility at its physical origin, rather than inferring it indirectly from reconstructed outputs. The two perspectives are therefore complementary but not equivalent.

\section{Design implications for physical imaging systems}
\label{sec:design_implications}

\medskip
\noindent The operator-based framework developed here has direct implications for the design and evaluation of physical imaging systems.

\medskip
\noindent Source design influences both operator entropy and information capacity. Increasing coherence or spectral bandwidth may increase entropy by redistributing energy across modes, but capacity improves only if those modes remain above the noise threshold. Detector improvements that reduce noise can increase capacity without altering entropy by lowering the effective threshold.

\medskip
\noindent Sampling strategy plays a decisive role in irreversibility. Undersampling introduces hard elimination of modes, increasing irreversibility even when resolution metrics appear acceptable. Progressive attenuation, by contrast, may preserve recoverability across a wider range of modes despite reducing apparent sharpness.

\medskip
\noindent These distinctions suggest that imaging system optimisation should be treated as a capacity-management problem rather than solely a resolution or signal-to-noise problem. Design choices should be evaluated based on how they affect operator spectra, effective rank, and irreversibility, not just downstream image appearance.

\medskip
\noindent The framework also provides a principled basis for comparing alternative system architectures and for identifying which physical stages dominate information loss. This enables targeted design interventions that improve recoverability rather than cosmetic image quality.

\medskip
\noindent In practical terms, the operator-based measures introduced here can be evaluated from measured or modelled system transfer functions or from calibrated forward models. For linear shift-invariant stages, the singular values are given (up to discretisation) by the sampled magnitude of the optical transfer function (OTF) or modulation transfer function (MTF); for general systems, they can be obtained by forming the discretised forward operator used in reconstruction and computing its singular spectrum (or a randomised/truncated approximation when $N$ is large). They therefore provide actionable diagnostics for system design, comparison, and optimisation, rather than abstract descriptors.

\section{Discussion}

\medskip
\noindent The operator-based formulation developed here has direct consequences for how imaging systems are analysed, compared, and designed. By separating intrinsic, operator-induced information loss from downstream noise, priors, and reconstruction choices, the framework clarifies which limitations are fundamentally irreversible and which are contingent on implementation. Improvements in detectors, algorithms, or exposure cannot recover information eliminated by hard spectral truncation, strong attenuation, or non-invertible physical transforms. Conversely, operator spectra provide a principled basis for comparing imaging modalities and system configurations on equal footing, independent of task-specific reconstructions or assumed signal statistics. This shifts system assessment away from image-space performance metrics toward operator-level descriptions of information flow.

\paragraph{Irreversible information loss at the operator level}
A key consequence of the operator-based formulation is that it identifies classes of information
loss that are fundamentally unrecoverable. When a measurement operator contracts, annihilates, or
collapses degrees of freedom, the corresponding information is removed at the level of the physical measurement process itself. No reconstruction algorithm can uniquely recover degrees of freedom that lie outside the range of the measurement operator from the measured data alone. The value of $\mathcal{I}_\epsilon(\mathcal{O})$ depends on $\epsilon$ because it counts both hard loss (numerically $\sigma_i \le \delta$ on the chosen discretisation) and soft loss ($\delta<\sigma_i<\epsilon$).

\medskip
\noindent In contrast, \emph{hard loss} is an $\epsilon$-independent property of $\mathcal{O}$: if the discretised forward operator has a non-trivial numerical nullspace (i.e.\ $\mathrm{rank}_\delta(\mathcal{O})<N$ on the chosen discretisation), then those nullspace modes are unrecoverable from data for every $\epsilon>0$ and, equivalently, for the limiting case $\epsilon\to 0^+$.

\medskip
\noindent This establishes a clear distinction between losses that are algorithmic and losses that are physical. Algorithmic losses arise from suboptimal inversion, discretisation, or numerical approximation and may, in principle, be mitigated. Operator-level losses arise from the structure of the measurement and persist regardless of downstream processing. The present framework makes this distinction explicit and provides a principled way to identify when further optimisation is informationally futile.

\medskip
\noindent The operator-based framework introduced in this paper provides a general way to quantify information flow in imaging. By focusing on the spectral structure of the operators that represent physical transformations, we can separate information-preserving behaviour from information-destroying behaviour in a consistent manner. The three measures introduced here: operator entropy; operator information capacity; irreversibility, capture different aspects of this behaviour.

\medskip
\noindent Operator entropy describes how an operator redistributes energy across its singular spectrum. It is sensitive to the diversity of modes that the operator retains, but does not specify which modes can be reliably recovered. Capacity addresses this by identifying the number of modes that remain above a noise-dependent threshold. Irreversibility characterises how much information is lost to suppression or elimination of modes and cannot be restored by any inverse or pseudo-inverse operator.

\medskip
\noindent Together, these measures provide a unified description of imaging transformations that is independent of the underlying modality. Attenuation, blur, sampling, and noise act in different ways but can be analysed within the same framework. The examples in Section~6 show that the operator viewpoint captures the essential differences between these mechanisms without requiring modality-specific assumptions.

\paragraph{Relationship to classical information theory}
It is important to emphasise that the operator-based framework introduced here is not intended to
replace classical information theory. Shannon entropy, mutual information, and related quantities
remain essential tools for analysing data, noise, and statistical dependence. The contribution of
the present work is to introduce an additional layer of description, operating at the level of
physical measurement operators rather than data alone. While equivalent descriptions may be
constructed in extended probabilistic spaces, such constructions require information about the
measurement operator that is not contained in data statistics alone.

\medskip
\noindent In this layered view, classical information theory characterises how information is distributed within observed data, while operator-based information theory characterises how information can, or cannot, propagate through a measurement process. The two perspectives address different questions and are most powerful when used together. This separation clarifies long-standing ambiguities in imaging and inverse problems, particularly regarding the limits of reconstruction and the meaning
of irreversibility.

\medskip
\noindent This framework also clarifies the distinction between physical and computational limits. Physical operators determine the maximum achievable information capacity and the irreversibility introduced by the imaging chain. Reconstruction algorithms cannot increase the effective rank of the measurement operator or reduce its irreversibility at a fixed operating threshold; they can only redistribute, stabilise, or bias the information contained within the surviving subspace. This has direct implications for the assessment of imaging systems and for the design of reconstruction methods.

\medskip
\noindent Claims of information recovery should therefore be interpreted as improvements in estimation within an operator-limited information budget, not as recovery of physically eliminated modes.

\medskip
\noindent The results developed here form the basis for further work on information geometry, spatiotemporal information budgets, nonlinear channels, and the information properties of reconstruction algorithms. These directions extend the operator-based approach and allow complete imaging pipelines to be analysed in a consistent information-theoretic framework.

\section{Conclusion}

\medskip
\noindent This paper introduced an operator-based formulation of information theory for imaging. By expressing physical imaging systems as composed operators acting on function spaces, it becomes possible to define measures of entropy, information capacity, and irreversibility that quantify how these transformations affect the recoverability of information.

\medskip
\noindent The operator entropy measures the distribution of energy across the singular spectrum. The information capacity describes the number of modes that remain above a noise threshold. The irreversibility index captures the information lost through the elimination or suppression of modes. These quantities apply to linear, nonlinear, and stochastic operators and are independent of the specific imaging modality.

\medskip
\noindent The framework provides a general approach for analysing physical limits in imaging, and distinguishes clearly between information that is lost through physical transformations and information that remains available for reconstruction. It also establishes the mathematical foundation for a broader information-theoretic treatment of imaging pipelines developed in subsequent work.

\medskip
\noindent The operator-based viewpoint offers a coherent route for comparing imaging configurations, characterising the effects of attenuation, blur, and sampling, and evaluating the limits imposed by noise and irreversibility. It provides a basis for the development of information geometry, capacity bounds for nonlinear channels, and analysis of the information properties of reconstruction algorithms. The results presented here form the starting point for a general theory of information
flow in imaging.

\ack{This work emerged through sustained analysis and synthesis of existing results in imaging science, inverse problems, and operator theory. The author acknowledges this literature for providing the conceptual context in which the present framework was developed.}

\funding{This research received no specific grant from any funding agency in the public, commercial, or not-for-profit sectors.}

\roles{Charles Wood: conceptualisation, formal analysis, methodology, writing of original draft, and subsequent review and editing.}

\data{Data sharing is not applicable to this article as no new data were generated or analysed in this study.}

\suppdata{No supplementary material is associated with this article}

\bibliographystyle{iopart-num}  

\bibliography {references}       

\end{document}